\documentclass[sigconf,authorversion]{acmart}

\usepackage{makecell}
\usepackage{caption}
\usepackage{booktabs}
\usepackage{subcaption}
\usepackage{hyperref}
\usepackage{macros}

\usepackage{csquotes}
\usepackage{multirow}

\usepackage{listings}

\usepackage{cuted}

\AtBeginDocument{%
  \providecommand\BibTeX{{%
    \normalfont B\kern-0.5em{\scshape i\kern-0.25em b}\kern-0.8em\TeX}}}

\copyrightyear{2021}
\acmYear{2021}
\setcopyright{acmlicensed}\acmConference[SIGCSE '21]{Proceedings of the 52nd ACM Technical Symposium on Computer Science Education}{March 13--20, 2021}{Virtual Event, USA}
\acmBooktitle{Proceedings of the 52nd ACM Technical Symposium on Computer Science Education (SIGCSE '21), March 13--20, 2021, Virtual Event, USA}
\acmPrice{15.00}
\acmDOI{10.1145/3408877.3432465}
\acmISBN{978-1-4503-8062-1/21/03}

\begin{document}
\fancyhead{}

\settopmatter{printfolios=false, printacmref=false}

\title[Using Comics to Introduce and Reinforce Programming Concepts in CS1]{Using Comics to Introduce and Reinforce \\Programming Concepts in CS1}

\author{Sangho Suh$^1$, Celine Latulipe$^2$, Ken Jen Lee$^1$, Bernadette Cheng$^1$, Edith Law$^1$}
\affiliation{
  \institution{$^1$University of Waterloo,  $^2$University of Manitoba, Canada}
  }
\email{sangho.suh@uwaterloo.ca, celine.latulipe@umanitoba.ca}
\email{{kenjen.lee, bhcheng, edith.law}@uwaterloo.ca}

\begin{abstract}
Recent work investigated the potential of comics to support the teaching and learning of programming concepts and suggested several ways \textit{coding strips}, a form of comic strip with its corresponding code, can be used. Building on this work, we tested the recommended use cases of \textit{coding strip} in an undergraduate introductory computer science course at a large comprehensive university. At the end of the course, we surveyed students to assess their experience and found they benefited in various ways. Our work contributes a demonstration of the various ways comics can be used in introductory CS courses and an initial understanding of benefits and challenges with using comics in computing education gleaned from an analysis of students' survey responses and code submissions.
\end{abstract}

\renewcommand{\shortauthors}{Suh et al.}



\maketitle

{\fontsize{8pt}{8pt} \selectfont
\textbf{ACM Reference Format:}\\
Sangho Suh, Celine Latulipe, Ken Jen Lee, Bernadette Cheng, Edith Law. 2021. Using Comics to Introduce and Reinforce Programming Concepts in CS1.  In {\it Proceedings of the 52nd ACM Technical Symposium on Computer Science Education (SIGCSE’21), March 13--20, 2021, Virtual Event, USA. } ACM, New York, NY, USA, 7 pages. https://doi.org/10.1145/3408877.3432465 }

\section{Introduction}
Critical goals in CS1 courses include motivating students and increasing their confidence in programming~\cite{kinnunen2006students, petersen2016revisiting}. This remains a challenge, however, due to several hurdles: programming requires students to work with unfamiliar conventions and syntax, and learn to trace the sequence of execution steps in a program---a difficult task for novice learners without visualization tools, training, or both~\cite{guo2013online, xie2018explicit}.
To address these challenges, recent work~\cite{suh2019using, suh2020promoting, suh2020coding} looked at comics, as it is a familiar medium capable of effectively communicating a sequence of events, with growing evidence and arguments in support of its use in education~\cite{yang2016comicsbelong}. Research in dual coding theory~\cite{clark1991dual, delp1996communicating, wang2014effects} provides theoretical and empirical support for comics. The theory posits that we process information using verbal and visual channels; thus, presenting information in both codes increases the chance of our remembering the information compared to when it is presented in only one code~\cite{paivio1975coding, delp1996communicating}. Since comics combine words and images to create dual-coded information~\cite{cohn2016multimodal, web:pond2018}, it is a potentially ideal ``dual-coding medium''~\cite{web:next-classroom-superhero, mcclanahan2019suite}. Recent work found evidence of dual coding effects in comics, further strengthening this idea~\cite{aleixo2017memory}.

\begin{figure}[htbp!]
    \centering
    \includegraphics[trim=0cm 0.5cm 0cm 0cm, clip=true, width=0.47\textwidth]{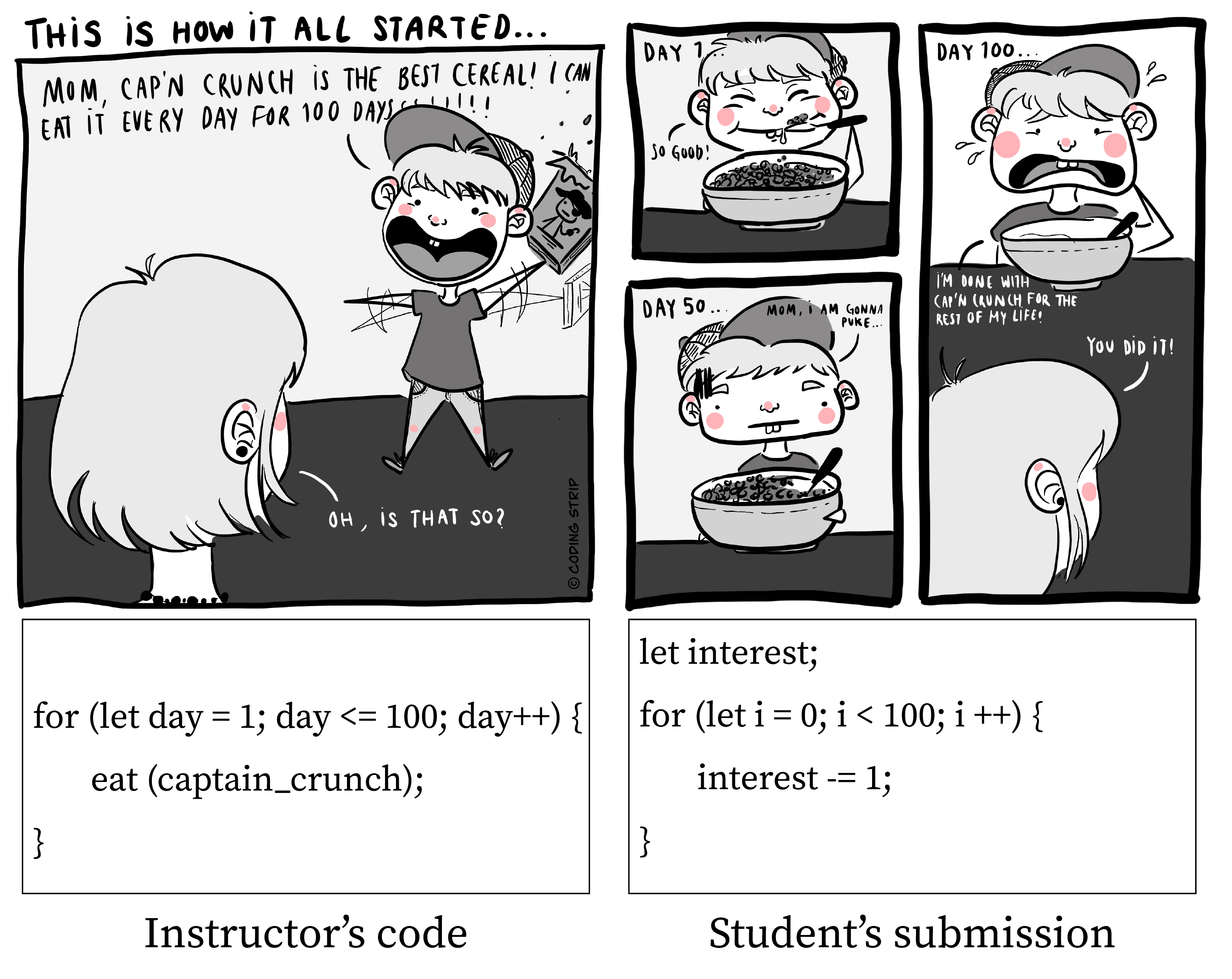}
    \vspace{-0.03in}
    \caption{Coding strip example on for loop used for one of the use cases: writing code from comics (\textbf{UC4}). Students were asked to translate the comics into code. A sample submission shows how comics can be interpreted in multiple ways.}
    \label{fig:code_writing-for_loop}
\end{figure}

Building on this work, we proposed \textit{coding strip}, a form of comic strip with corresponding code (e.g., Fig.~\ref{fig:code_writing-for_loop}), as a tool for teaching and learning programming concepts~\cite{suh2020coding}. While we found several ways students and instructors wanted to use \textit{coding strip}, we did not examine those use cases with students in a classroom setting. As a result, we do not know what the associated benefits and challenges are. Moreover, without a report detailing their administration, it is unclear how instructors can use \textit{coding strip}. Thus we administered these use cases in an undergraduate CS1 course and surveyed students to understand the benefits and challenges. This experience report fills this gap and makes the following contributions:

\begin{itemize}
    \item description of four use cases of coding strips in a CS1 course,
    \item analysis of perceived usefulness of comics and use cases, 
    \item summary of benefits and challenges with using \it{coding strip}.
\end{itemize}

\section{Methods}
\textit{Coding strip} was used to teach programming concepts in a CS1 course in four ways. In this section, we describe the course, the four use cases of \textit{coding strip}, and the post-semester survey.

\subsection{Course \& Student Information}
The study was conducted in a first-year computer science course (N=49) at the University of Waterloo. The course is designed primarily for students who do not major in computer science (e.g., students in Arts). Seventy percent of students in this course were in the Digital Arts Program, and for them the course was required. The course followed the {\it creative coding} approach~\cite{peppler2005creative}, where students learn computer programming by manipulating media (e.g., graphics, sound) and creating interactive graphics. Students used p5.js, a popular Javascript library for creative coding. Therefore students learned and programmed using Javascript syntax and conventions.

\subsection{Use Cases}
While the four use cases of \textit{coding strip} we tested are not an exhaustive list of all potential use cases (\textbf{UC}s), they represent common teaching tasks relevant for most, if not all, programming courses.

\subsubsection{\textbf{UC1. Introduce Concept}} 
\begin{figure}[htbp!]
    \centering
    \includegraphics[trim=0cm 0cm 0cm 0cm, clip=true, width=0.48\textwidth]{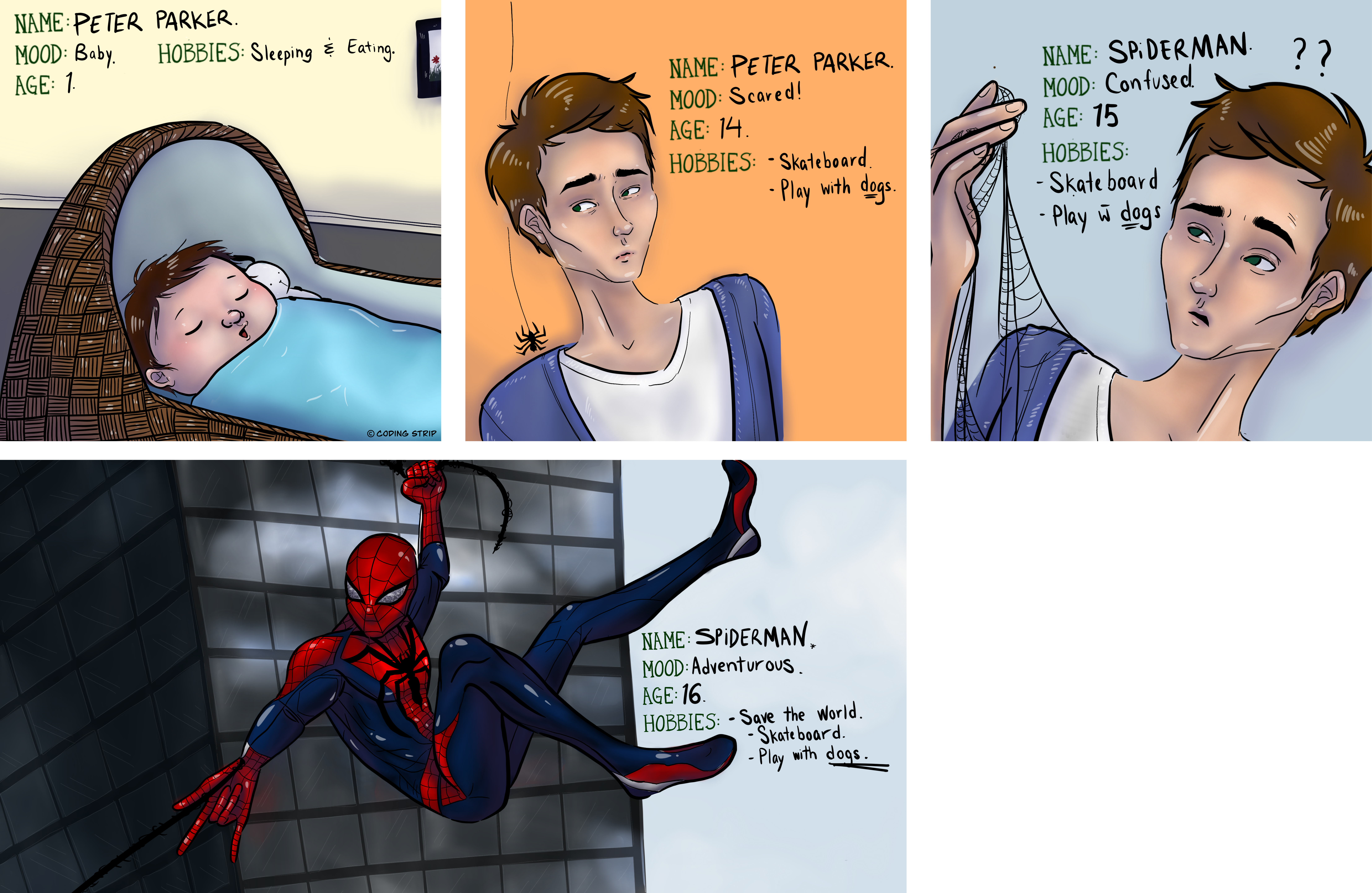}
    \caption{A Spiderman comic used to introduce the idea that values in variables change (\textbf{UC1}). Here, variables are \textit{name}, \textit{mood}, \textit{age}, and \textit{hobbies}.}
    \label{fig:introducing_variable}
\end{figure}

\begin{figure*}[th!]
    \centering
    \begin{subfigure}[t]{0.326\textwidth}
        \includegraphics[trim=1cm 0cm 1cm 0cm, clip=true, width=\textwidth]{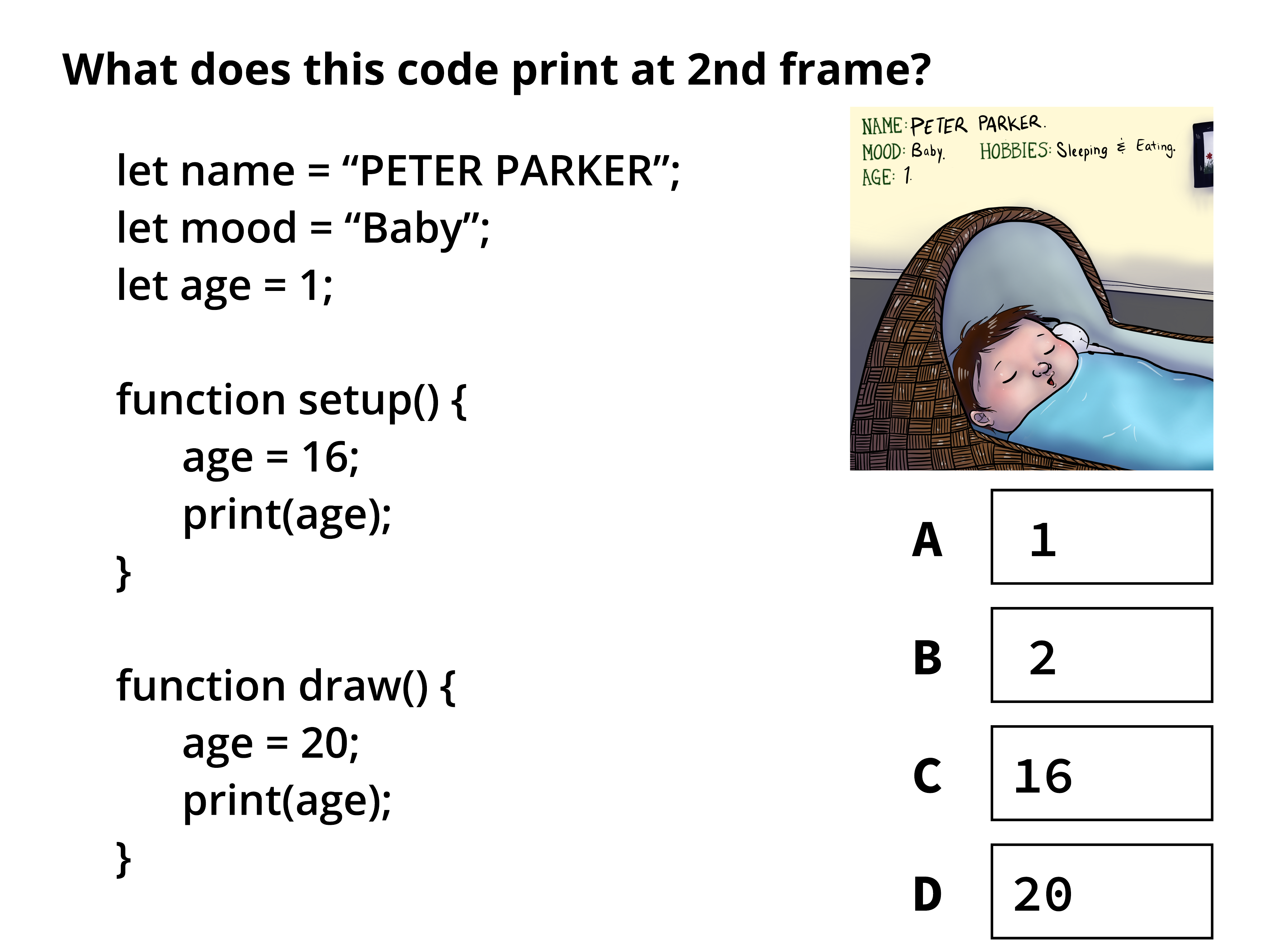}
        \caption{\textbf{UC3-variable1}}
        \label{fig:clicker-variable}
    \end{subfigure}
    \begin{subfigure}[t]{0.326\textwidth}
        \includegraphics[trim=1cm 0cm 1cm 0cm, clip=true, width=\textwidth]{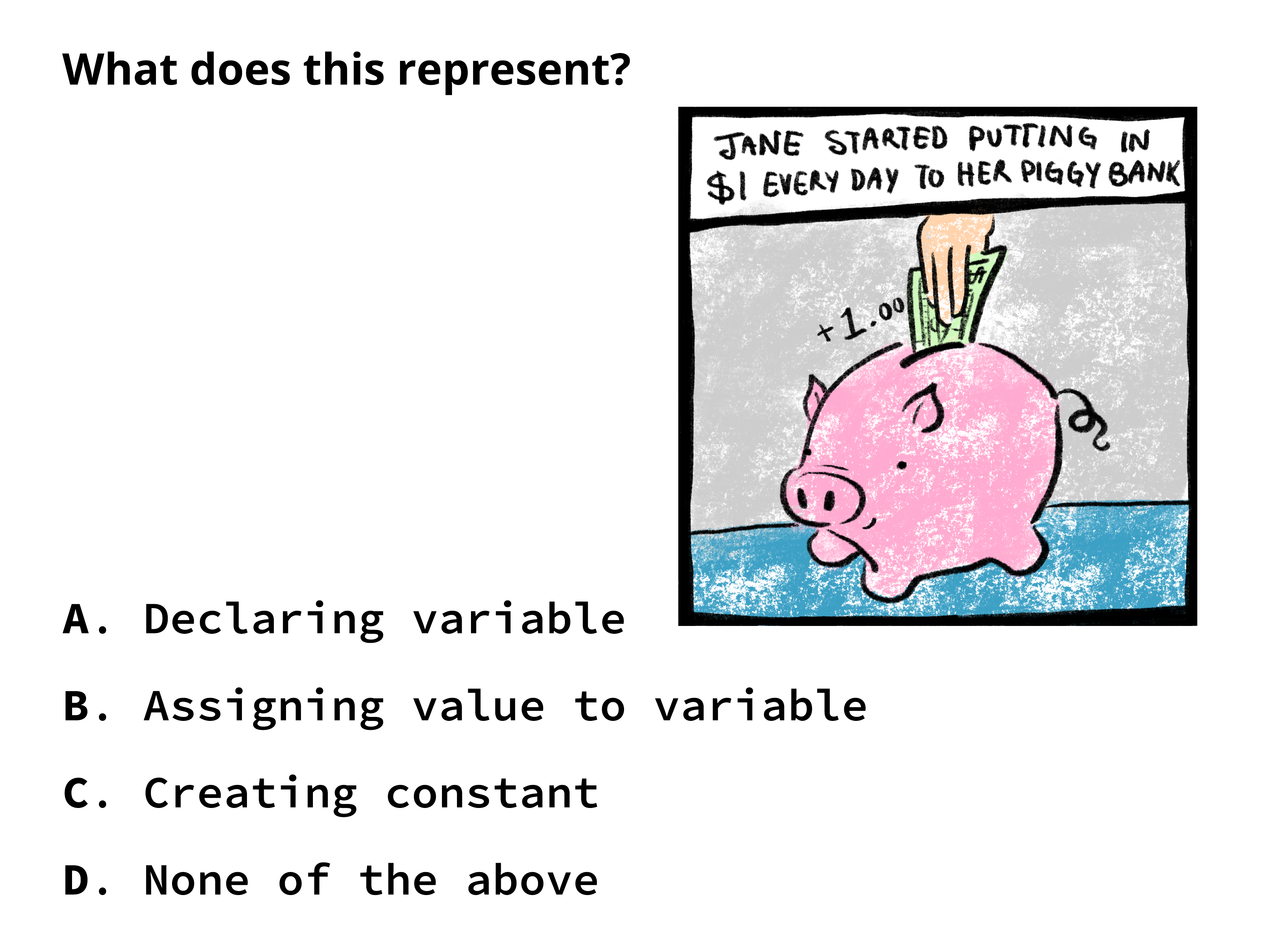}
        \caption{\textbf{UC3-variable2}}
        \label{fig:clicker-variable_assignment}
    \end{subfigure}
    \begin{subfigure}[t]{0.326\textwidth}
        \includegraphics[trim=1cm 0cm 1cm 0cm, clip=true, width=\textwidth]{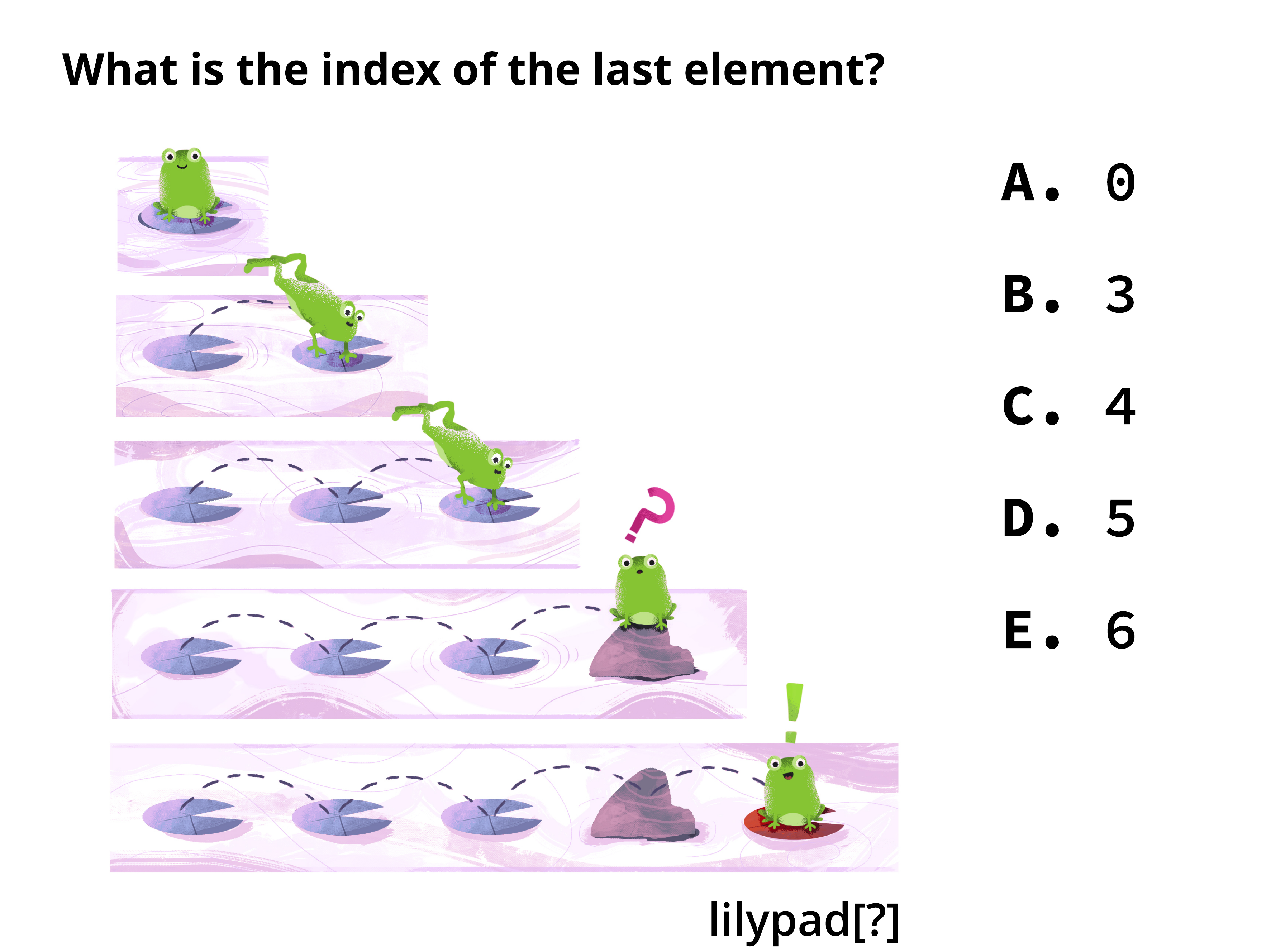}
        \caption{\textbf{UC3-array}}
        \label{fig:clicker-array}
    \end{subfigure}
    \caption{Examples of reviewing with clicker questions (\textbf{UC3})}
    \vspace{-0.05in}
    \label{fig:reviewing_with_comic}
\end{figure*}

The Spiderman comic shown in Fig.~\ref{fig:introducing_variable} was used in week 3 of the course when the concept \textit{variable} was being introduced for the first time. The course instructor displayed a slide with the comic before the start of the class for students arriving early. Once the lecture began, students learned that variables store values and that associated values can change over time. The instructor then showed the Spiderman comic again relating it to how values associated with Peter's ``name'', ``mood'', ``age'', and ``hobbies'' change over time. For this use case, only the comic was used; its corresponding code was later used in a clicker question for review (Fig.~\ref{fig:clicker-variable}).

\begin{figure}[h!]
    \centering
    \begin{subfigure}[t]{0.48\textwidth}
        \includegraphics[trim=0cm 0.5cm 0cm 0cm, clip=true, width=\textwidth]{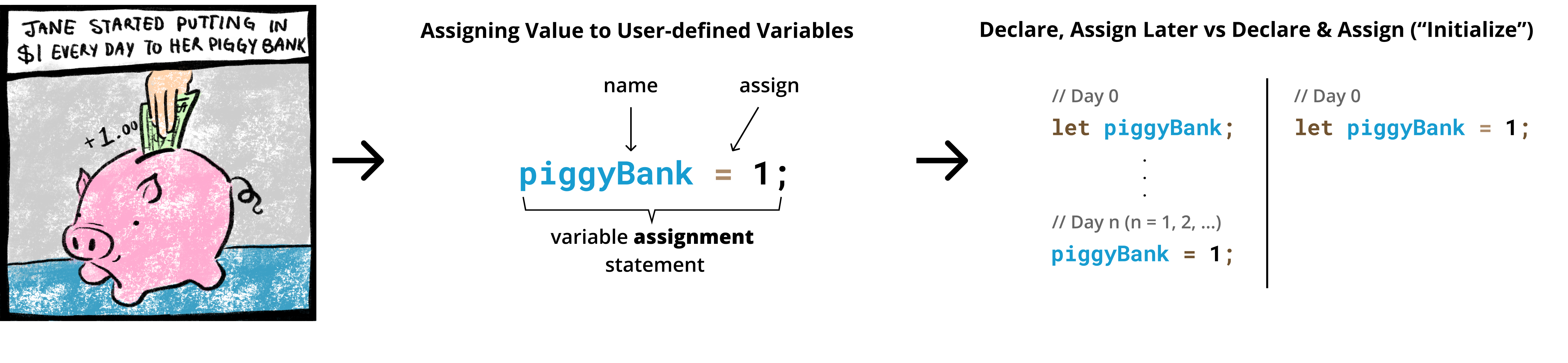}
        \caption{Variable assignment, declaration, initialization}
        \label{fig:introduce-variable_code}
    \end{subfigure}
    \begin{subfigure}[t]{0.48\textwidth}
        \includegraphics[trim=0cm 0cm 0cm 0cm, clip=true, width=\textwidth]{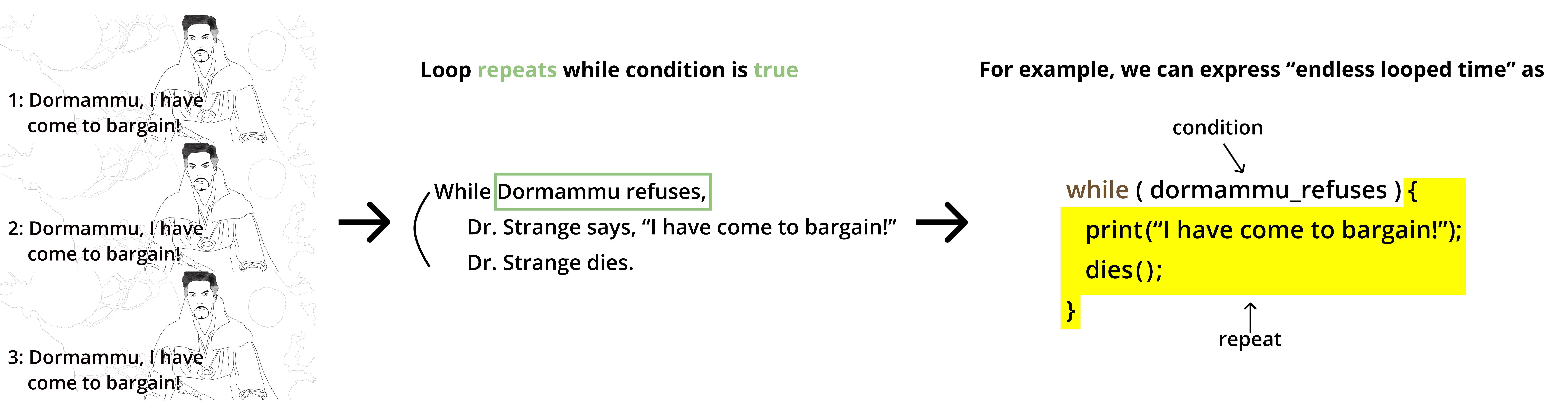}
        \caption{While loop}
        \label{fig:introduce-loop_code}
    \end{subfigure}    
    \begin{subfigure}[t]{0.48\textwidth}
        \includegraphics[trim=0cm 0.2cm 0cm 0.9cm, clip=true, width=\textwidth]{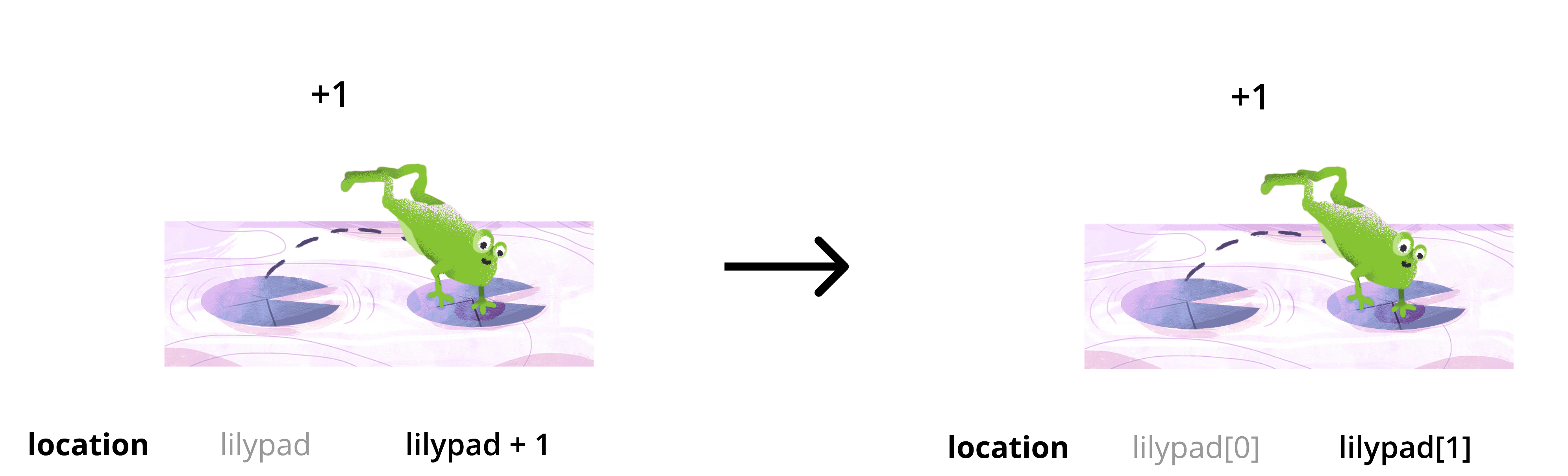}
        \caption{Array}
        \label{fig:introduce-array_code}
    \end{subfigure}
    \caption{Sequence of lecture slides used to introduce code with comics (\textbf{UC2})}
    \vspace{-0.15in}
    \label{fig:introducing_code}
\end{figure}

\subsubsection{\textbf{UC2. Introduce Code}}
Inspired by prior work~\cite{suh2019using, suh2020how}, comics were used to scaffold code expression in several ways, as shown in Fig.~\ref{fig:introducing_code}. Fig.~\ref{fig:introduce-variable_code} shows a sequence of three lecture slides which begins with a comic strip that portrays a character putting money into a piggy bank for the concept \textit{variable assignment}, followed by its corresponding code, and a slide with its related concepts \textit{variable declaration} and \textit{initialization} in the same piggy bank context; in other words, it (1) begins with an intuitive, relatable abstraction, (2) connects it to its code expression, and (3) explains the concept by comparing and contrasting with related concepts. Fig.~\ref{fig:introduce-loop_code} shows another sequence of slides which progresses from comic to English and to code; students were shown a clip from the movie \textit{Dr. Strange}~\cite{strange2016} in which Dr. Strange traps a villain Dormammu in an endless looped time; then, the instructor showed three panels of an image (first slide of the sequence) from the clip to highlight the repeating sequence; this scenario was then presented in English (with indentations to mimic the code syntax) and finally its corresponding code. Fig.~\ref{fig:introduce-array_code} shows a progression from a slide with comic \& English to a slide with comic \& code. In this sequence, the goal was to introduce the code expression for the array and highlight how array indices start at 0, not 1. As shown, students were first shown the comic with a frog jumping from one lily pad to another. In the following slide, the expressions ``lilypad'' and ``lilypad + 1'' changed to ``lilypad[0]'' and ``lilypad[1]'' to show the corresponding code expressions in array syntax. The ``+1'' above the frog was used to provide an intuitive explanation for why array indices start at 0 and not 1. This is something novice learners often struggle to grasp because they think indices represent ordinal numbers (e.g., index 1 points to 1st element); they do not realize that indices represent offsets (i.e., index 1 points to 2nd element, index 1 means 1 position away from the 1st element).

\subsubsection{\textbf{UC3. Review Concepts and Code}}
After introducing concepts and code expressions with comics, students can review them using the same comic, its code, or context. Fig.~\ref{fig:reviewing_with_comic} shows three clicker questions used to review concepts and code that were presented in  Fig.~\ref{fig:introducing_variable} and \ref{fig:introducing_code}. Fig.~\ref{fig:clicker-variable} required students to track changing values in a variable ``age'' which was previously shown in the Spiderman comic (Fig.~\ref{fig:introducing_variable}). Fig.~\ref{fig:clicker-variable_assignment} asked students to select an appropriate abstraction (``assigning value to variable'') for the piggy bank illustration from Fig.~\ref{fig:introduce-variable_code}. Fig.~\ref{fig:clicker-array} was used to test whether students understood that array indexing starts from 0 as they were taught with Fig.~\ref{fig:introduce-array_code}.

\subsubsection{\textbf{UC4. Write Code from Comics}}
Another use case suggested in prior work~\cite{suh2020coding} was coding exercises, that is, have students translate comics into code. Fig.~\ref{fig:code_writing-for_loop} and \ref{fig:code_writing} show three coding strips used for the exercises. During the lecture, students were shown the comics without the code and were asked to submit their code on Socrative, a web-based response system~\cite{web:socrative}. Since it is unreasonable to expect students to submit the exact same code as the instructor, students were promised full participation marks for simply submitting. After students submitted, the instructor showed the list of submissions and his code, and used submissions that reflect students' unique interpretation of comics to highlight that coding is a tool for creative expression.

\vspace{-0.05in}
\subsection{Survey}
At the end of the semester, we asked students to complete a survey to evaluate their experience with each use case and the idea of using comics (we used the word \textit{comics} instead of \textit{coding strip} in the survey to avoid unnecessarily overloading students with new terminology). While one of the authors was the instructor of the course, we followed the guidelines from the university’s ethics committee to ensure we did not exert undue influence on the students to give us permission to use their data or to answer with any bias. Specifically, a research assistant unaffiliated with the course administered the Google form survey via the course's online platform and collected the survey responses. Students were assured the instructor would not know who had permitted the use of their data until after their final course grades were submitted.

In order to help students remember and differentiate the four different use cases, we included a page in the survey with the description and lecture slides used for each use case (Fig.~\ref{fig:code_writing-for_loop},~\ref{fig:introducing_variable},~\ref{fig:introducing_code},~\ref{fig:reviewing_with_comic},~\ref{fig:code_writing} were used), before the questions about each use case. The survey began with some demographic and programming attitude questions, and then included questions about the use of comics, which progressed from questions about the use of comics in general and then to questions specific to each of the four use cases. At the end, they were also asked to specify whether they thought other computer science instructors should use comics. Scale-based questions about the comics used a 7-point Likert scale where 1 indicated ``Really Disliked'' and 7 indicated ``Really Liked''.

\vspace{-0.05in}
\section{Results}
We present the analysis of survey responses about all use cases and analysis of code submissions for use case 4 (\textbf{UC4}). Specifically, we describe what students liked and disliked about each use case, as well as why students do or do not recommend using comics in other computing courses, in order to present the benefits and challenges with using comics. To ensure anonymity, we refer to students as S1...S41.

\vspace{-0.05in}
\subsection{Demographics}
Of the 49 students enrolled in the course, 42 students completed the survey and 41 students consented to the use of their data for research. Hence, our analysis is based on these 41 students (15 M, 26 F; Arts: 28; Science: 9; Health Science: 4). Most students were undergraduate students in their first or second year (18 First, 15 Second, 4 Third, 3 Fourth), with one student in a post-degree program. The majority of the students were taking the course to meet degree requirements (28), while the rest (13) were taking the course out of interest in learning programming. Half of the students were retaking the course (21 Retake, 20 First). In terms of their programming experience before the course, most students had limited or no experience (15 No Experience, 18 Several Hours or Days, 8 Several Weeks or Months). Students were evenly split in terms of their interest in learning programming at the beginning of the course (15 Not Interested, 16 Interested to Certain Degree, 10 Highly Interested). At the beginning of the course, more than half of the students (25) perceived learning programming as ``difficult'', while 14 saw it as ``manageable'' and 2 saw it as ``easy''.

\vspace{-0.05in}
\subsection{Analysis of Each Use Case}

\begin{figure}[htbp!]
    \centering
    \includegraphics[trim=0cm 0cm 0cm 0cm, clip=true, width=0.48\textwidth]{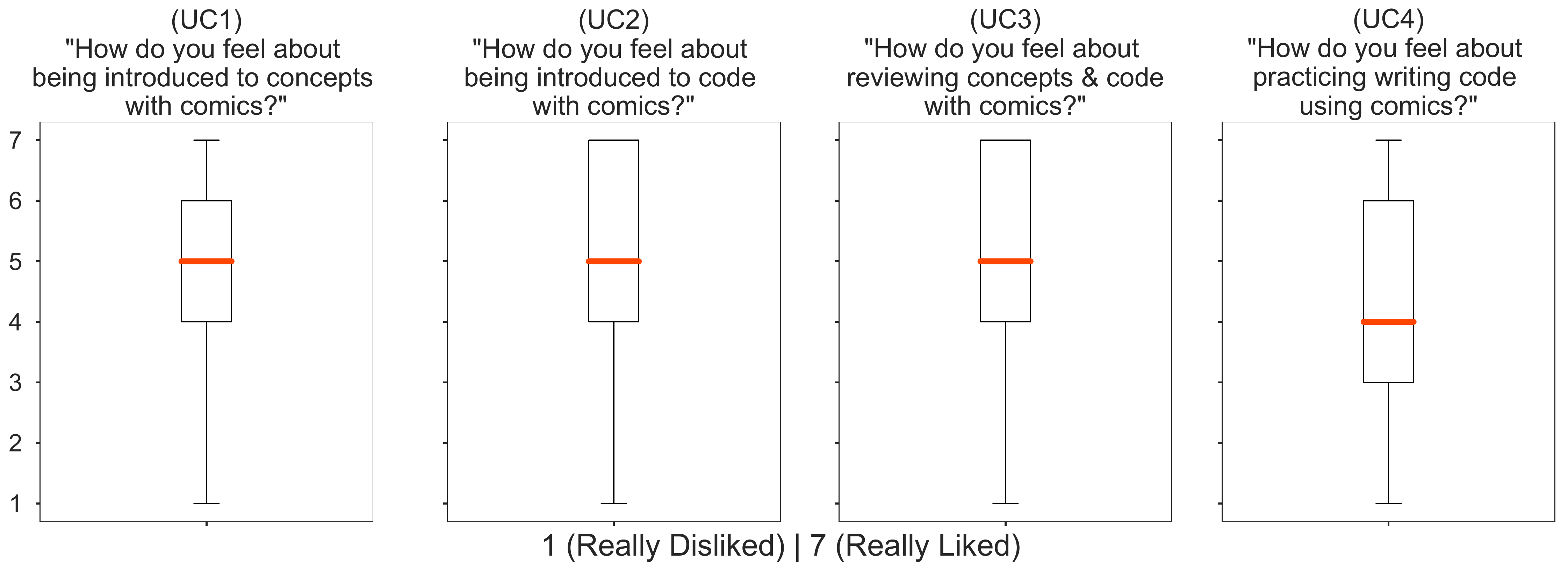}
    \vspace{-0.25in}
    \caption{Students' assessment of four use cases}
    \vspace{-0.15in}
    \label{fig:boxplot}
\end{figure}

\subsubsection{\textbf{UC1. Introduce Concept}} Students generally liked being introduced to concepts with comics (M=4.9/7, 95\%CI=[4.3,5.4]). Thirty-five students (85\%) who rated it positively (scores of 4-7) explained that they liked it because it was not only ``fun'' (S1, S29), ``engaging'' (S32), ``related to something[, such as superheroes, they] like'' (S8), ``an interesting way to learn [the concepts]'' (S33, S37), but also ``piqued [their] interest for the concept'' (S9), and made programming less ``scary'' (S40) and the ``material less dry'' (S32).

Students also mentioned that comics helped them better understand the concepts (S7, S28, S33, S35), as they explain ``why'' (S17), help them ``visualize the concept'' (S13, S15, S41), ``simplify the tricky concepts into a more digestible format'' (S26), provide ``good analogy'' (S22) and ``metaphor'' (S2), and give ``another point of view'' (S28) to help make sense of the concept. The sequential nature of comics was also helpful in understanding the procedural aspect of the concepts:

\begin{displayquote}
``[Comics] gave a logical explanation to the concepts applied. A lot of the time, we don't know what the program is doing; the comics made a logical sequence of concepts that made it easier to learn.'' (S19)
\end{displayquote}

Another important benefit was that comics helped students remember and easily recall the concepts; S27 and S5 mentioned that they recalled the comics ``during the midterm'' (S27) and whenever they needed to remember, for instance, ``what loop does'' (S5).

Six students (15\%) who rated this use case negatively (scores of 1-3) offered two major explanations. One was that the comics were confusing (S3, S16, S20); S25 said he had difficulty understanding ``how [the comics and concepts] correlate.'' The other argument was that the method is ``not applicable to all students'' (S34), which coincided with S27's comment on his preference for learning with analogy alone to learning with comics.

Students also shared several ideas on how we could improve this use case. S30 observed that since ``it may be harder to understand [the comics] if the student isn't familiar with the [concept already]'', we may want to use comics some time after the student has learned the concept. Several students (S9, S20, S23) also suggested that comics may be more useful when introducing ``more complex topics, like loops and arrays'' (S9) and ``concepts which need step-by-step explanations (ex: loops)'' (S23). This seemed to align with S31, who found the frog comic (Fig.~\ref{fig:introduce-array_code})---which shows a frog leaping from one lily pad to another in step-by-step manner to illustrate looping through an array---more useful than the variable comic (Fig.~\ref{fig:introducing_variable}).

\subsubsection{\textbf{UC2. Introduce Code}}
Students also liked being introduced to code with comics (M=5/7, 95\%CI=[4.4,5.6]). Many of the benefits reported for this use case were similar to those for \textbf{UC1}: it was ``engaging'' (S7), ``fun'' (S31), and a nice way to introduce ``the code in a more interesting way'' (S14, S26). Several students also reported that introducing the code with comics made the code ``easier to remember'' (S5, 18) compared to being introduced to code with text only. S40 also contrasted it with the text-only approach to point out that learning code with comics allowed her to focus on understanding the code instead of merely memorizing it, which she resorted to whenever code was presented in text only.

Students (S1, S35, S40) reasoned that comics made code ``easier to understand'' (S1) because they provide visual structure (S39) and show ``the logic behind code'' (S22). S19 explained how comics helped her make sense of loops: ``loops are hard to understand because [you need to make sense of] the exit point and amount of times it should run, but when Dormammu finally says stop, it made sense way easier than learning it from scratch.'' Finally, some students mentioned that being introduced to code in this manner ``relieved some anxiety'' (S2) and helped them develop a positive attitude towards learning programming (S26).

Like \textbf{UC1}, a few students mostly found comics ``confusing'' (S15) and ``not useful'' (S17). S3 and S24 thought it was ``unnecessary'' because they assumed ``all [students already] understood the concept of a loop.'' Since most students, however, commented on their usefulness, it seems that it was a welcome intervention for most students 
other than a few who already understood the code.

\subsubsection{\textbf{UC3. Review Concepts and Code}}
Students also enjoyed reviewing the learned concepts and code with comics (M=4.9/7, 95\%CI=[4.4,5.5]). They said it helped them better understand and remember the review content (S8, S31) and that it is ``fun'' (S11, S21) and ``a good way to refresh memories'' (S20, S22), as it makes them ``pay attention'' (S1). S14 shared that while ``[he] was confused by some comics'', he still liked it ``because [reviewing with comics] was a good checkpoint for [him] to determine if [he] can apply the concept.'' S19 described how helpful one of our review questions (Fig.~\ref{fig:clicker-array}) was for her understanding of array:

\begin{displayquote} 
``Absolutely brilliant. Using a leap frog going through an array because we don't know how to grasp the concept that the program goes from 0 to 1 to 2 in a set of array ... this HELPED A LOT.'' (S19)
\end{displayquote}

While it needs further testing, we also found that students generally performed better on clicker questions when they referenced comics (Fig.~\ref{fig:reviewing_with_comic}). The average percentage of correct answers for non-comic based clicker questions (60\%) was lower than that of the comic-based clicker questions (74\% for \textbf{UC3-variable1}, 67\% for \textbf{UC3-variable2}, 86\% for \textbf{UC3-array}). We also directly compared \textbf{UC3-variable1} with its isomorphic (i.e., corresponding in format/task) question, which was administered in the next class without comics references (i.e., ``let x = 1;'' instead of ``let age = 1;'' and no comics in the slide), and found that only 34\% of the students responded correctly compared to 74\% for \textbf{UC3-variable1}. We did not administer isomorphic questions for the other two.

\begin{figure}[b!]
    \centering
    \includegraphics[trim=0cm 0cm 0cm 0cm, clip=true, width=0.48\textwidth]{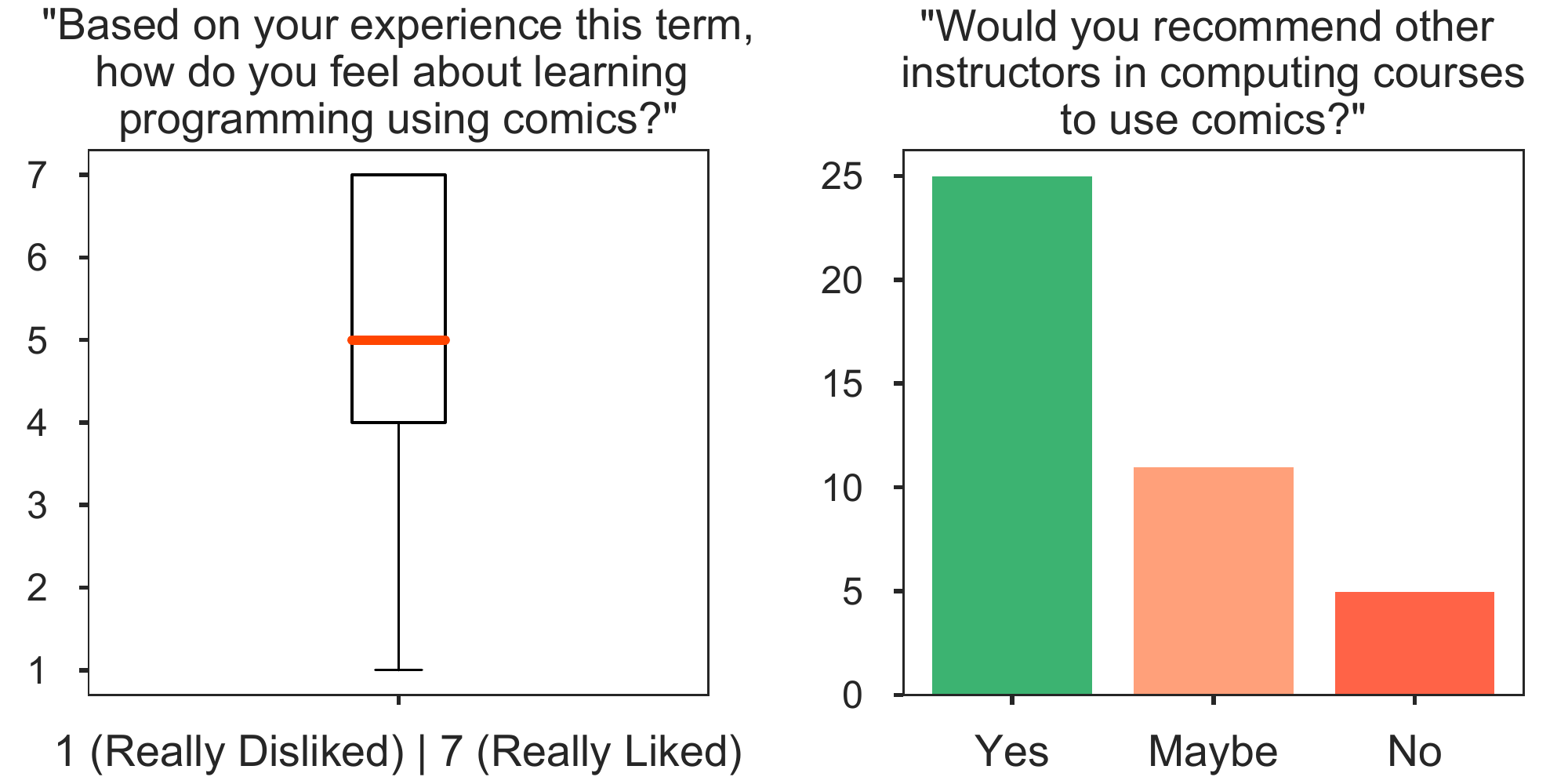}
    \caption{Students' assessment of comics in general}
    \label{fig:comic_overall}
\end{figure}

While most students rated this use case positively, some students reminded us that a few factors could potentially undermine students' experience with this use case. S15 noted that he was able to enjoy reviewing with comics because he already knew the concept well enough and thus did not find the comics confusing. S40 who rated this use case highly (6/7) warned us that since this use case affects students' grades, we may want to ensure that the comics are clear. While the clicker question grades served more or less as a participation grade and represented a small portion of the overall grades (they were 5\% of the total grade and the best 75\% of clicker grades were used), a few students were careful about fully supporting it as they felt some students got certain questions wrong ``simply because the [comic] was confusing, rather than them not understanding the topic'' (S3).

\begin{figure}[thbp!]
    \begin{subfigure}[t]{0.438\textwidth}
        \includegraphics[trim=0cm 0cm 0cm 0cm, clip=true, width=\textwidth]{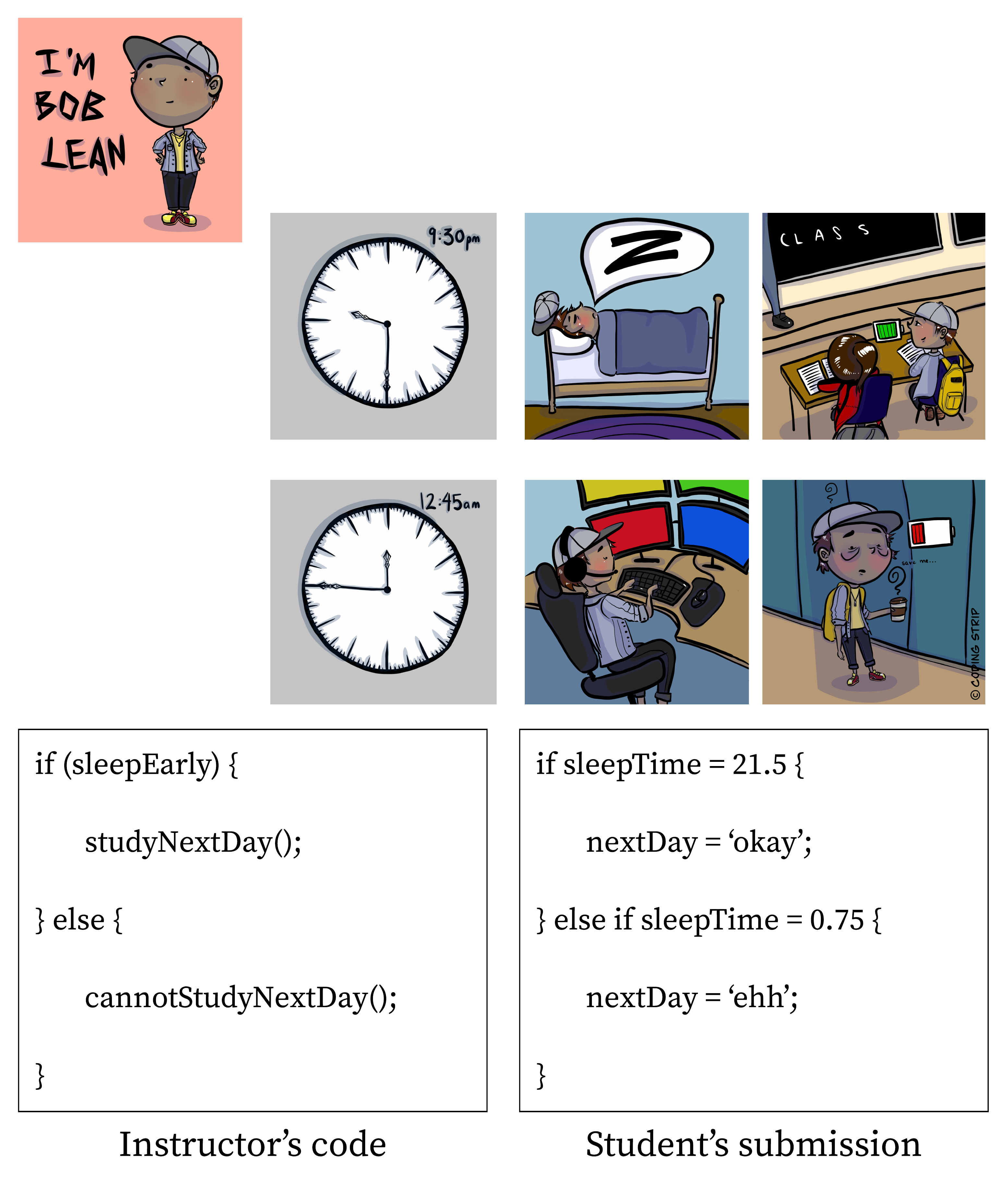}
        \caption{\textbf{Boolean}}
        \label{fig:code_writing-boolean}
    \end{subfigure}
    \centering
    \begin{subfigure}[t]{0.438\textwidth}
        \includegraphics[trim=0cm 0cm 0cm 0cm, clip=true, width=\textwidth]{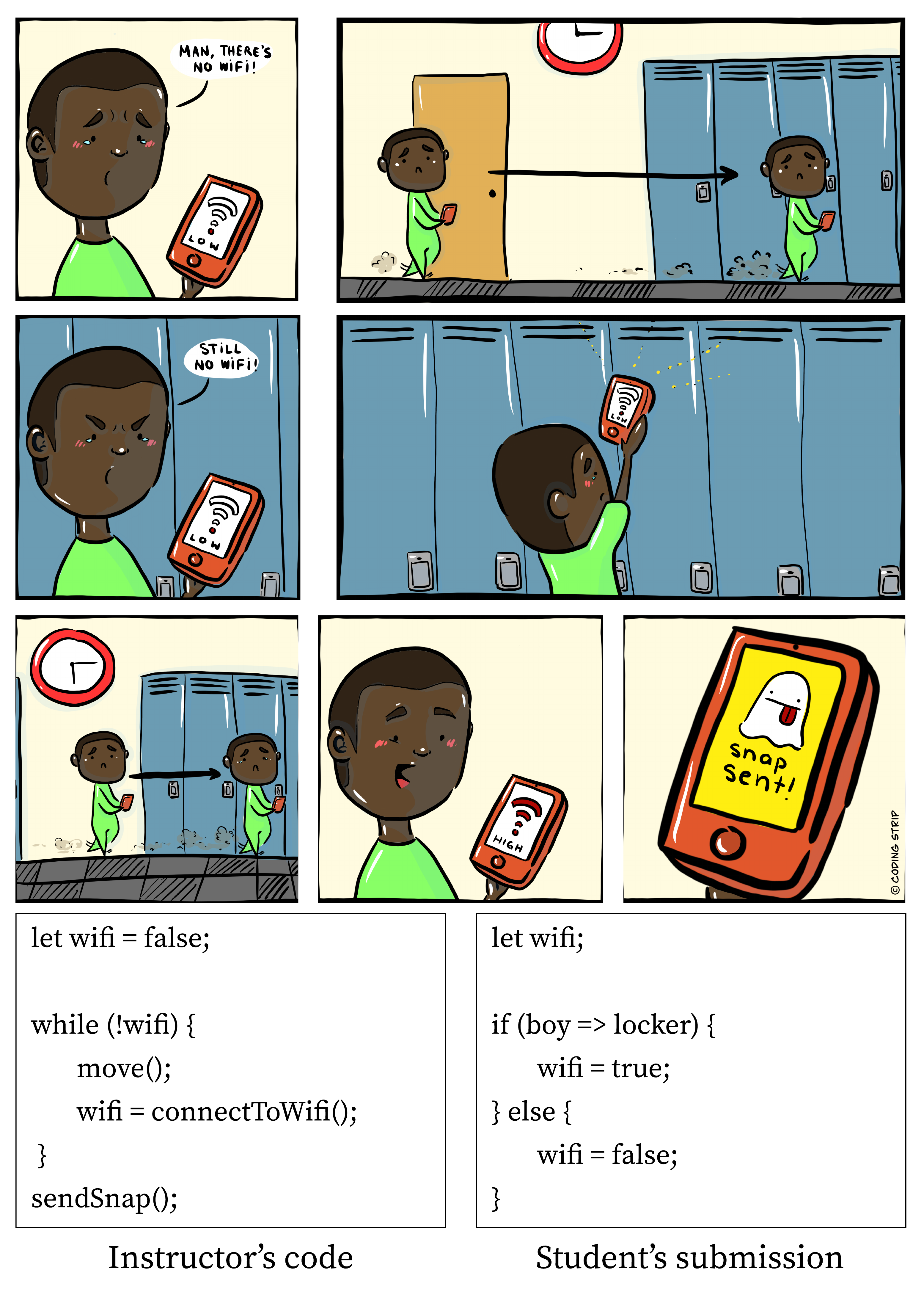}
        \caption{\textbf{While loop}}
        \label{fig:code_writing-while_loop}
    \end{subfigure}
    \captionsetup{justification=centering}
    \caption{Two of the three coding strips used for writing code from comics (\textbf{UC4}). One other coding strip is Fig.~\ref{fig:code_writing-for_loop}}
    \label{fig:code_writing}
\end{figure}

\subsubsection{\textbf{UC4. Write Code from Comics}}
Compared to other use cases, students were not as positive about the idea of practicing writing code from comics (M=4.2/7, 95\%CI=[3.6,4.8]). Many students found the exercise ``difficult'' (S11, S14, S18, S20, S25) because they had to write code and interpret comics. This is understandable since students can find writing code from scratch challenging and interpreting comics difficult due to the task's open-ended nature. Since these resulted in a less positive experience and confusion among some students (S3, S5, S24), this exercise would benefit from clear guidelines with examples of code submissions. On the other hand, there were also students who understood it after the first time (S12), were interested in ``learn[ing] more'' (S8), found it ``fun'' (S18, S31) and ``very useful to practice and go over the concepts in a limited time'' (S26). Some students felt the exercise ``[made] programming seem less intimidating'' (S39) and more interesting:

\begin{displayquote}
``I have always tried to find the `right' answer because I have been educated that there is only one right answer and I need to find it. So when I saw that [comic] in class, it seemed like a big difficulty and I didn't know what to do. However, this made me more interested in programming after realizing that in programming, there is no right answer and the result depends on what I'm creating and expressing.'' (S40)
\end{displayquote}

This comment on how the exercise changed S40's view on programming was an unexpected, yet exciting, finding. Upon analyzing students' code submissions, we found many interesting interpretations of comics, some of which are shown in Fig.~\ref{fig:code_writing-for_loop} and \ref{fig:code_writing}. 

\vspace{-0.03in}
\subsection{Analysis of Overall Experience} 
Students, in general, were highly positive (M=5.2/7, 95\%CI=[4.7,5.7]) about the overall idea of learning programming using comics. When students were asked at the end of the survey whether they would recommend that other instructors teaching computing courses use comics, more than half of the students were in support, and very few were against it (25 Yes, 11 Maybe, 5 No). Students were asked to elaborate on their response and we present the analysis of their answers below.

\subsubsection{Reasons for Recommending}
More than 60\% of students recommended the use of comics in other computing courses for a variety of reasons. Many students recited the benefits from previous sections, including the comics being ``fun'' (S21, S35), ``engaging'' (S18, S32), ``motivating'' (S31, S35), helpful in ``remembering'' the concepts and code (S22, S27). S7, who was taking the course for the second time, compared his experience from the previous term to stress that comics helped him understand the materials better. Students also mentioned that comics help ``break down big coding ideas'' (S31) and ``give a visual representation of what is happening'' (S20) and that it is an ``appealing way to learn'' (S8) for students who prefer learning with visuals (S6, S12, S15, S39). Last but not least, they also spoke of the positive impact it has on the classroom atmosphere by making lectures more engaging (S1, S20, S21, S29).

Students were divided, however, on the extent to which comics may be useful in other computing courses (Figure~\ref{fig:comic_overall}). S32 said, ``it definitely works for the lower level CS classes. Maybe not for the CS major classes, though.'' On the other hand, S22 believed it could be used for learning any programming language. S8 noted that since the ``majority of people like comics'', it would be a welcome intervention in any course. S4, who had prior programming experience, speculated that while he did not need the comics, other students with a little programming background would like learning with comics.

\subsubsection{Reasons for Reservation}
More than a quarter of students had reservations about comics being used in other CS courses. S27 said it would be okay ``if they don't solely rely on it as a teaching method.'' Some were hesitant because they felt that comics' usefulness depends on a number of factors, such as ``[the clarity of the] the [comics'] link to code'' (S24), ``[its] content'', ``the student's learning style'' (S10), ``how [much they] enjoy/understand comics'' (S1), and which use cases are used (S15, S23, S36).

\subsubsection{Reasons for Not Recommending}
Five students who did not recommend the use of comics in computing courses explained that they found comics ``confusing'' (S3), not ``useful'' (S17), not suiting their ``learning style'' (S25) and ``hard to see from the back [of the classroom]'' (S34). S38 thought that ``because all computing courses cater to different needs [of the students] it is not for everyone.''

\vspace{-0.03in}
\subsection{Analysis of Demographics}
We examined whether students' assessment of the overall use of comics and each use case varied by demographic attributes---gender, major, the reason for taking the course, prior experience with the course, prior programming experience, prior interest in learning programming. We found no statistically significant differences between the groups within these categories. This is noteworthy because it suggests that students' experiences with coding strips are similar, regardless of their backgrounds. However, our sample size was small and differences across these demographics may emerge in future work with larger samples.

\vspace{-0.02in}
\section{Discussion}
Students generally enjoyed learning with comics and experienced various benefits, some of which were related to dual coding effects, such as being able to ``remember'' and ``understand'' the concepts and code more easily~\cite{paivio1975coding, kounios1994concreteness, aleixo2017memory}. Although our study does not contribute any measurement of learning impact, it still provides valuable findings to help facilitate the use of \textit{coding strip}. Specifically, it contributes an understanding of how coding strips can be used to support teaching in an undergraduate CS1 course, how students perceive them, and what needs to be improved.

Our study also found that while the idea of learning style has been dismissed due to lack of evidence~\cite{pashler2008learning, rohrer2012learning, howard2014neuroscience}, numerous students still held this misconception and cited their learning style as the reason for liking or recommending the comics. Likewise, this was also the reason some students disliked or hesitated to recommend comics: they thought non-visual learners do not benefit from comics as it is ``visual-based.'' In addition to stigma associated with comic books, we find that this misconception about learning style may be another challenge that needs to be addressed since it seems to significantly impact students' perception and acceptance.

As we move forward, the following issues need to be addressed. First, while the use cases were generally well-received, the coding exercise (\textbf{UC4}) needs to be improved. For instance, it needs a clear process to guide students and a variety of examples to help them feel comfortable with the idea that code can have multiple interpretations. Also, a number of students attributed the ambiguity of the comics causing confusion as the reason for disliking the use cases or not recommending comics. While a few students described what caused the confusion (e.g., lack of clear link between the comics and code), there is still a lack of clear knowledge on what (e.g., particular design aspect of the comics) made them confusing. Finally, our results imply that students' perception of \textit{coding strips} can depend on the timing of when the comics are introduced; more investigations are needed to understand these more nuanced effects.

While we used comics throughout the course, they were not used for all concepts and code. The concepts spanned multiple levels of difficulty (i.e., variable as ``easy'', loop as ``medium'', and array as ``hard''). However, it is unclear if and how this may have impacted students' assessment. Furthermore, the students in the class were non-CS majors, with most of the survey responses (68\%) coming from students in the Digital Arts Program. Thus we need to investigate whether CS majors experience the same benefits. While we explored several ways \textit{coding strips} can be used in this study, we did not explore what aspects made certain coding strips more or less useful or confusing for learners. As future work, we can analyze specific coding strips' styles and their learning effects in order to develop design guidelines for \textit{coding strips}.
Finally, we need to find out in what alternative format or ways the instructors should use coding strips to accommodate blind or visually impaired students.

\vspace{-0.03in}
\section{Conclusion}
In this work, we tested four use cases of \textit{coding strips} and contribute an experience report to share how they were used, how each use case was perceived by students, as well as the benefits and challenges associated with each use case. While the four use cases we deployed are not exhaustive, they address basic teaching tasks and we believe this contributes a useful report for instructors interested in using coding strips to introduce programming concepts and code. While more work is needed to understand the most effective ways to use coding strips, our work contributes an essential first step towards their use in computing education.

\section{Acknowledgement}
This research was funded by Learning Innovation and Technology Enhancement (LITE) Grant at the University of Waterloo. We would also like to thank the students for their participation and reviewers for their feedback and suggestions.

\bibliographystyle{ACM-Reference-Format}
\bibliography{main}


\begin{thebibliography}{25}


\ifx \showCODEN    \undefined \def \showCODEN     #1{\unskip}     \fi
\ifx \showDOI      \undefined \def \showDOI       #1{#1}\fi
\ifx \showISBNx    \undefined \def \showISBNx     #1{\unskip}     \fi
\ifx \showISBNxiii \undefined \def \showISBNxiii  #1{\unskip}     \fi
\ifx \showISSN     \undefined \def \showISSN      #1{\unskip}     \fi
\ifx \showLCCN     \undefined \def \showLCCN      #1{\unskip}     \fi
\ifx \shownote     \undefined \def \shownote      #1{#1}          \fi
\ifx \showarticletitle \undefined \def \showarticletitle #1{#1}   \fi
\ifx \showURL      \undefined \def \showURL       {\relax}        \fi
\providecommand\bibfield[2]{#2}
\providecommand\bibinfo[2]{#2}
\providecommand\natexlab[1]{#1}
\providecommand\showeprint[2][]{arXiv:#2}

\bibitem[\protect\citeauthoryear{Aleixo}{Aleixo}{2017}]%
        {web:next-classroom-superhero}
\bibfield{author}{\bibinfo{person}{Paul~A Aleixo}.}
  \bibinfo{year}{2017}\natexlab{}.
\newblock \bibinfo{title}{How the humble comic book could become the next
  classroom superhero}.
\newblock
  \bibinfo{howpublished}{\url{https://theconversation.com/how-the-humble-comic-book-could-become-the-next-classroom-superhero-73486}}.
\newblock
\newblock
\shownote{Accessed: 2020-06-17.}


\bibitem[\protect\citeauthoryear{Aleixo and Sumner}{Aleixo and Sumner}{2017}]%
        {aleixo2017memory}
\bibfield{author}{\bibinfo{person}{Paul~A Aleixo} {and}
  \bibinfo{person}{Krystina Sumner}.} \bibinfo{year}{2017}\natexlab{}.
\newblock \showarticletitle{Memory for biopsychology material presented in
  comic book format}.
\newblock \bibinfo{journal}{\emph{Journal of Graphic Novels and Comics}}
  \bibinfo{volume}{8}, \bibinfo{number}{1} (\bibinfo{year}{2017}),
  \bibinfo{pages}{79--88}.
\newblock


\bibitem[\protect\citeauthoryear{Clark and Paivio}{Clark and Paivio}{1991}]%
        {clark1991dual}
\bibfield{author}{\bibinfo{person}{James~M Clark} {and} \bibinfo{person}{Allan
  Paivio}.} \bibinfo{year}{1991}\natexlab{}.
\newblock \showarticletitle{Dual coding theory and education}.
\newblock \bibinfo{journal}{\emph{Educational psychology review}}
  \bibinfo{volume}{3}, \bibinfo{number}{3} (\bibinfo{year}{1991}),
  \bibinfo{pages}{149--210}.
\newblock


\bibitem[\protect\citeauthoryear{Cohn}{Cohn}{2016}]%
        {cohn2016multimodal}
\bibfield{author}{\bibinfo{person}{Neil Cohn}.}
  \bibinfo{year}{2016}\natexlab{}.
\newblock \showarticletitle{A multimodal parallel architecture: A cognitive
  framework for multimodal interactions}.
\newblock \bibinfo{journal}{\emph{Cognition}}  \bibinfo{volume}{146}
  (\bibinfo{year}{2016}), \bibinfo{pages}{304--323}.
\newblock


\bibitem[\protect\citeauthoryear{Delp and Jones}{Delp and Jones}{1996}]%
        {delp1996communicating}
\bibfield{author}{\bibinfo{person}{Chris Delp} {and} \bibinfo{person}{Jeffrey
  Jones}.} \bibinfo{year}{1996}\natexlab{}.
\newblock \showarticletitle{Communicating information to patients: the use of
  cartoon illustrations to improve comprehension of instructions}.
\newblock \bibinfo{journal}{\emph{Academic Emergency Medicine}}
  \bibinfo{volume}{3}, \bibinfo{number}{3} (\bibinfo{year}{1996}),
  \bibinfo{pages}{264--270}.
\newblock


\bibitem[\protect\citeauthoryear{Derrickson}{Derrickson}{2016}]%
        {strange2016}
\bibfield{author}{\bibinfo{person}{Scott Derrickson}.} \bibinfo{year}{Doctor
  Strange. Marvel Studios, Hollywood, 2016}\natexlab{}.
\newblock
\newblock


\bibitem[\protect\citeauthoryear{Guo}{Guo}{2013}]%
        {guo2013online}
\bibfield{author}{\bibinfo{person}{Philip~J Guo}.}
  \bibinfo{year}{2013}\natexlab{}.
\newblock \showarticletitle{Online python tutor: embeddable web-based program
  visualization for cs education}. In \bibinfo{booktitle}{\emph{Proceeding of
  the 44th ACM technical symposium on Computer science education}}. ACM,
  \bibinfo{pages}{579--584}.
\newblock


\bibitem[\protect\citeauthoryear{Howard-Jones}{Howard-Jones}{2014}]%
        {howard2014neuroscience}
\bibfield{author}{\bibinfo{person}{Paul~A Howard-Jones}.}
  \bibinfo{year}{2014}\natexlab{}.
\newblock \showarticletitle{Neuroscience and education: myths and messages}.
\newblock \bibinfo{journal}{\emph{Nature Reviews Neuroscience}}
  \bibinfo{volume}{15}, \bibinfo{number}{12} (\bibinfo{year}{2014}),
  \bibinfo{pages}{817--824}.
\newblock


\bibitem[\protect\citeauthoryear{Kinnunen and Malmi}{Kinnunen and
  Malmi}{2006}]%
        {kinnunen2006students}
\bibfield{author}{\bibinfo{person}{P{\"a}ivi Kinnunen} {and}
  \bibinfo{person}{Lauri Malmi}.} \bibinfo{year}{2006}\natexlab{}.
\newblock \showarticletitle{Why students drop out CS1 course?}. In
  \bibinfo{booktitle}{\emph{Proceedings of the second international workshop on
  Computing education research}}. \bibinfo{pages}{97--108}.
\newblock


\bibitem[\protect\citeauthoryear{Kounios and Holcomb}{Kounios and
  Holcomb}{1994}]%
        {kounios1994concreteness}
\bibfield{author}{\bibinfo{person}{John Kounios} {and}
  \bibinfo{person}{Phillip~J Holcomb}.} \bibinfo{year}{1994}\natexlab{}.
\newblock \showarticletitle{Concreteness effects in semantic processing: ERP
  evidence supporting dual-coding theory.}
\newblock \bibinfo{journal}{\emph{Journal of Experimental Psychology: Learning,
  Memory, and Cognition}} \bibinfo{volume}{20}, \bibinfo{number}{4}
  (\bibinfo{year}{1994}), \bibinfo{pages}{804}.
\newblock


\bibitem[\protect\citeauthoryear{McClanahan and Nottingham}{McClanahan and
  Nottingham}{2019}]%
        {mcclanahan2019suite}
\bibfield{author}{\bibinfo{person}{Barbara~J McClanahan} {and}
  \bibinfo{person}{Maribeth Nottingham}.} \bibinfo{year}{2019}\natexlab{}.
\newblock \showarticletitle{A suite of strategies for navigating graphic
  novels: A dual coding approach}.
\newblock \bibinfo{journal}{\emph{The Reading Teacher}} \bibinfo{volume}{73},
  \bibinfo{number}{1} (\bibinfo{year}{2019}), \bibinfo{pages}{39--50}.
\newblock


\bibitem[\protect\citeauthoryear{Paivio}{Paivio}{1975}]%
        {paivio1975coding}
\bibfield{author}{\bibinfo{person}{Allan Paivio}.}
  \bibinfo{year}{1975}\natexlab{}.
\newblock \showarticletitle{Coding distinctions and repetition effects in
  memory}.
\newblock \bibinfo{journal}{\emph{The psychology of learning and motivation}}
  \bibinfo{volume}{9} (\bibinfo{year}{1975}), \bibinfo{pages}{179--214}.
\newblock


\bibitem[\protect\citeauthoryear{Pashler, McDaniel, Rohrer, and Bjork}{Pashler
  et~al\mbox{.}}{2008}]%
        {pashler2008learning}
\bibfield{author}{\bibinfo{person}{Harold Pashler}, \bibinfo{person}{Mark
  McDaniel}, \bibinfo{person}{Doug Rohrer}, {and} \bibinfo{person}{Robert
  Bjork}.} \bibinfo{year}{2008}\natexlab{}.
\newblock \showarticletitle{Learning styles: Concepts and evidence}.
\newblock \bibinfo{journal}{\emph{Psychological science in the public
  interest}} \bibinfo{volume}{9}, \bibinfo{number}{3} (\bibinfo{year}{2008}),
  \bibinfo{pages}{105--119}.
\newblock


\bibitem[\protect\citeauthoryear{Peppler and Kafai}{Peppler and Kafai}{2005}]%
        {peppler2005creative}
\bibfield{author}{\bibinfo{person}{K Peppler} {and} \bibinfo{person}{Y Kafai}.}
  \bibinfo{year}{2005}\natexlab{}.
\newblock \showarticletitle{Creative coding: Programming for personal
  expression}.
\newblock \bibinfo{journal}{\emph{Retrieved August}} \bibinfo{volume}{30},
  \bibinfo{number}{2008} (\bibinfo{year}{2005}), \bibinfo{pages}{314}.
\newblock


\bibitem[\protect\citeauthoryear{Petersen, Craig, Campbell, and
  Tafliovich}{Petersen et~al\mbox{.}}{2016}]%
        {petersen2016revisiting}
\bibfield{author}{\bibinfo{person}{Andrew Petersen}, \bibinfo{person}{Michelle
  Craig}, \bibinfo{person}{Jennifer Campbell}, {and} \bibinfo{person}{Anya
  Tafliovich}.} \bibinfo{year}{2016}\natexlab{}.
\newblock \showarticletitle{Revisiting why students drop CS1}. In
  \bibinfo{booktitle}{\emph{Proceedings of the 16th Koli Calling International
  Conference on Computing Education Research}}. \bibinfo{pages}{71--80}.
\newblock


\bibitem[\protect\citeauthoryear{Pond}{Pond}{2017}]%
        {web:pond2018}
\bibfield{author}{\bibinfo{person}{Shayna Pond}.}
  \bibinfo{year}{2017}\natexlab{}.
\newblock \bibinfo{title}{Comics and Dual Coding Theory}.
\newblock
  \bibinfo{howpublished}{\url{http://shaynapond.oucreate.com/blog/uncategorized/comics-and-dual-coding-theory/}}.
\newblock
\newblock
\shownote{Accessed: 2020-03-11.}


\bibitem[\protect\citeauthoryear{Rohrer and Pashler}{Rohrer and
  Pashler}{2012}]%
        {rohrer2012learning}
\bibfield{author}{\bibinfo{person}{Doug Rohrer} {and} \bibinfo{person}{Harold
  Pashler}.} \bibinfo{year}{2012}\natexlab{}.
\newblock \showarticletitle{Learning Styles: Where's the Evidence?.}
\newblock \bibinfo{journal}{\emph{Online Submission}} \bibinfo{volume}{46},
  \bibinfo{number}{7} (\bibinfo{year}{2012}), \bibinfo{pages}{634--635}.
\newblock


\bibitem[\protect\citeauthoryear{Socrative}{Socrative}{2020}]%
        {web:socrative}
\bibfield{author}{\bibinfo{person}{Socrative}.}
  \bibinfo{year}{2020}\natexlab{}.
\newblock \bibinfo{howpublished}{\url{https://www.socrative.com/}}.
\newblock
\newblock
\shownote{Accessed: 2020-04-18.}


\bibitem[\protect\citeauthoryear{Suh}{Suh}{2019}]%
        {suh2019using}
\bibfield{author}{\bibinfo{person}{Sangho Suh}.}
  \bibinfo{year}{2019}\natexlab{}.
\newblock \showarticletitle{Using Concreteness Fading to Model \& Design
  Learning Process}. In \bibinfo{booktitle}{\emph{Proceedings of the 2019 ACM
  Conference on International Computing Education Research}}.
  \bibinfo{pages}{353--354}.
\newblock


\bibitem[\protect\citeauthoryear{Suh}{Suh}{2020}]%
        {suh2020promoting}
\bibfield{author}{\bibinfo{person}{Sangho Suh}.}
  \bibinfo{year}{2020}\natexlab{}.
\newblock \showarticletitle{Promoting Meaningful Learning by Supporting
  Interplay within Abstraction Ladder}. In \bibinfo{booktitle}{\emph{2020 IEEE
  Symposium on Visual Languages and Human-Centric Computing (VL/HCC)}}. IEEE.
\newblock


\bibitem[\protect\citeauthoryear{Suh, Lee, and Law}{Suh et~al\mbox{.}}{2020a}]%
        {suh2020how}
\bibfield{author}{\bibinfo{person}{Sangho Suh}, \bibinfo{person}{Martinet Lee},
  {and} \bibinfo{person}{Edith Law}.} \bibinfo{year}{2020}\natexlab{a}.
\newblock \showarticletitle{How Do We Design for Concreteness Fading? Survey,
  General Framework, and Design Dimensions}. In
  \bibinfo{booktitle}{\emph{Proceedings of the 19th ACM Conference on
  Interaction Design and Children}}.
\newblock


\bibitem[\protect\citeauthoryear{Suh, Lee, Xia, and Law}{Suh
  et~al\mbox{.}}{2020b}]%
        {suh2020coding}
\bibfield{author}{\bibinfo{person}{Sangho Suh}, \bibinfo{person}{Martinet Lee},
  \bibinfo{person}{Gracie Xia}, {and} \bibinfo{person}{Edith Law}.}
  \bibinfo{year}{2020}\natexlab{b}.
\newblock \showarticletitle{Coding Strip: A Pedagogical Tool for Teaching and
  Learning Programming Concepts through Comics}. In
  \bibinfo{booktitle}{\emph{2020 IEEE Symposium on Visual Languages and
  Human-Centric Computing (VL/HCC)}}. IEEE.
\newblock


\bibitem[\protect\citeauthoryear{Wang}{Wang}{2014}]%
        {wang2014effects}
\bibfield{author}{\bibinfo{person}{Ling Wang}.}
  \bibinfo{year}{2014}\natexlab{}.
\newblock \showarticletitle{The effects of single and dual coded multimedia
  instructional methods on Chinese character learning}.
\newblock \bibinfo{journal}{\emph{Chinese as a Second Language Research}}
  \bibinfo{volume}{3}, \bibinfo{number}{1} (\bibinfo{year}{2014}),
  \bibinfo{pages}{1--25}.
\newblock


\bibitem[\protect\citeauthoryear{Xie, Nelson, and Ko}{Xie
  et~al\mbox{.}}{2018}]%
        {xie2018explicit}
\bibfield{author}{\bibinfo{person}{Benjamin Xie}, \bibinfo{person}{Greg~L
  Nelson}, {and} \bibinfo{person}{Amy~J Ko}.} \bibinfo{year}{2018}\natexlab{}.
\newblock \showarticletitle{An explicit strategy to scaffold novice program
  tracing}. In \bibinfo{booktitle}{\emph{Proceedings of the 49th ACM Technical
  Symposium on Computer Science Education}}. \bibinfo{pages}{344--349}.
\newblock


\bibitem[\protect\citeauthoryear{Yang}{Yang}{2016}]%
        {yang2016comicsbelong}
\bibfield{author}{\bibinfo{person}{Gene Yang}.}
  \bibinfo{year}{2016}\natexlab{}.
\newblock \bibinfo{title}{Why comics belong in the classroom}.
\newblock
  \bibinfo{howpublished}{\url{https://www.youtube.com/watch?v=Oz4JqAJbxj0&feature=youtu.be}}.
\newblock
\newblock
\shownote{Accessed: 2020-02-10.}


\end{thebibliography}

\end{document}